\documentclass[twocolumn,floatfix,preprintnumbers,nofootinbib,superscriptaddress]{revtex4}

\usepackage{ulem}
\usepackage{bm}
\usepackage{times}
\usepackage{amssymb,amsbsy,amsmath,amsfonts}
\usepackage{graphicx}
\usepackage{float}
\usepackage{color}
\usepackage{morefloats}
\usepackage{rotating}
\usepackage{srcltx}
\usepackage{slashed}
\usepackage{subfigure}
\usepackage{multirow}
\usepackage{verbatim}
\usepackage{hyperref}
\usepackage{tabularx}

\usepackage[outdir=./]{epstopdf}

\usepackage{graphicx}
\usepackage{amsmath}
\usepackage{amsfonts}
\usepackage{amssymb}
\usepackage{color}
\usepackage{multirow}
\usepackage{mathrsfs}
\usepackage{gensymb}% \degree order needed

\usepackage{booktabs}  
\usepackage{threeparttable}

\usepackage{subfigure}

\newcommand\be{\begin{eqnarray}}
\newcommand\ee{\end{eqnarray}}

\begin{document}

\title{Nature of $X(3872)$ in $B^0 \to K^0 X(3872)$ and $B^+ \to K^+ X(3872)$ decays}

\author{Hao-Nan Wang}~\email{wanghaonan@m.scnu.edu.cn}
\affiliation{Guangdong Provincial Key Laboratory of Nuclear Science, Institute of Quantum Matter,
South China Normal University, Guangzhou 510006, China}
\affiliation{Institute of Modern Physics, Chinese Academy of Sciences, Lanzhou 730000, China}

\author{Li-Sheng Geng}~\email{lisheng.geng@buaa.edu.cn}
\affiliation{Peng Huanwu Collaborative Center for Research and Education, Beihang University, Beijing 100191, China}
\affiliation{School of Physics, Beihang University, Beijing 102206, China}
\affiliation{Beijing Key Laboratory of Advanced Nuclear Materials and Physics, Beihang University, Beijing, 102206, China}

\author{Qian Wang}~\email{qianwang@m.scnu.edu.cn}
\affiliation{Guangdong Provincial Key Laboratory of Nuclear Science, Institute of Quantum Matter,
South China Normal University, Guangzhou 510006, China}
\affiliation{Guangdong-Hong Kong Joint Laboratory of Quantum Matter, Southern Nuclear Science Computing Center,
South China Normal University, Guangzhou 510006, China}
\affiliation{Theoretical Physics Center for Science Facilities, Institute of High Energy Physics, Chinese Academy of Sciences,
Beijing 100049, China}

\author{Ju-Jun Xie}~\email{xiejujun@impcas.ac.cn}
\affiliation{Institute of Modern Physics, Chinese Academy of Sciences, Lanzhou 730000, China}
\affiliation{School of Nuclear Sciences and Technology, University of Chinese Academy of Sciences, Beijing 101408, China}

\begin{abstract}

We investigate the decays of $B^0 \to K^0 X(3872)$ and $B^+ \to K^+ X(3872)$ based on the picture where the  $X(3872)$ resonance is strongly coupled to the $D\bar{D}^* + c.c.$ channel. In addition to the decay mechanism where the $X(3872)$ resonance is formed from the $c\bar{c}$ pair hadronization with the short-distance interaction, we have also considered the $D\bar{D}^*$ rescattering diagrams in the long-distance scale, where $D$ and $\bar{D}^*$ are formed from $c$ and $\bar{c}$ separately. Because of the difference of the mass thresholds of charged and neutral $D\bar{D}^*$ channels, and the rather narrow width of the $X(3872)$ resonance, at the  $X(3872)$ mass, the loop functions of $D^0\bar{D}^{*0}$ and $D^+\bar{D}^{*-}$ are much different. Taking this difference into account, the ratio of $\mathcal{B}[B^0\to K^0X(3872)]/\mathcal{B}[B^+ \to K^+ X(3872)] \simeq 0.5$ can be naturally obtained. Based on this result, we also evaluate the decay widths of $B_s^0 \to \eta(\eta') X(3872)$. It is expected that future experimental measurements of these decays can be used to elucidate the nature of the $X(3872)$ resonance.

\end{abstract}

\maketitle

%%%%%%%%%%%%%%%%%%%%%%%%%%%%%%%%%%%%%%%%%%%%%%%%%%%%%%%%%%%%%%%%%%%%%%%%%%

\section{Introduction} \label{section:introduction}

The discovery of the $X(3872)$ [also known as $\chi_{c1}(3872)$] resonance by the Belle Collaboration in 2003~\cite{Belle:2003nnu} and confirmed later by other experiments~\cite{CDF:2003cab,D0:2004zmu,CDF:2006ocq,BaBar:2008qzi,CDF:2009nxk,LHCb:2011zzp,CMS:2013fpt,LHCb:2013kgk,LHCb:2015jfc}, ushered in a new era in hadron physics~\cite{Guo:2017jvc,Liu:2019zoy,Brambilla:2019esw}, which cannot be easily explained as a conventional charmonium state in quark models~\cite{Barnes:2003vb,Barnes:2005pb}. The isospin of $X(3872)$ is zero, and its quantum numbers $J^{PC} = 1^{++}$ were extracted by the LHCb Collaboration~\cite{LHCb:2013kgk,LHCb:2015jfc}. Particularly, the mass of $X(3872)$, $M_{X(3872)} = 3871.65 \pm 0.06$ MeV, is very close to the threshold of $D^0\bar{D}^{\ast0}$, $M_{D^0\bar{D}^{\ast0}} = 3871.69 \pm 0.11$ MeV, and its width is $\Gamma_{X(3872)} = 1.19 \pm 0.21$ MeV~\cite{ParticleDataGroup:2020ssz}, which is rather narrow. A precise determination of the mass of the $X(3872)$ resonance is crucial to understanding its nature~\cite{Guo:2019qcn}. Furthermore, the  branching fractions of  the $X(3872)$ resonance decaying into $J/\psi \rho^0$ and $J/\psi \omega$ are similar~\cite{LHCb:2022bly}. It is found that the contributions of the $\omega$ meson to the $X(3872) \to J/\psi \pi^+ \pi^-$ decays are sizeable~\cite{LHCb:2022bly,Wang:2022vjm}, which indicates that the $X(3872)$ resonance has sizeable coupling to the $J/\psi \omega$ channel. Recently, the  $e^+ e^- \to \omega X(3872)$ reaction was observed by the BESIII Collaboration~\cite{Yuan:2022lxf}.

About two decades since the discovery of the $X(3872)$ resonance, in spite of all the available experimental data measured by different collaborations~\cite{Belle:2003nnu,CDF:2003cab,D0:2004zmu,CDF:2006ocq,BaBar:2008qzi,CDF:2009nxk,LHCb:2011zzp,CMS:2013fpt,LHCb:2013kgk,LHCb:2015jfc}, the nature of the $X(3872)$ resonance is still unclear~\cite{Chen:2022asf,Dong:2021bvy,Guo:2019twa,Ali:2017jda,Olsen:2017bmm,Esposito:2016noz,Chen:2016qju,Lebed:2016hpi}. It has been interpreted as a tetraquark state of the diquark-antidiquark structure~\cite{Maiani:2004vq,Ebert:2005nc,Matheus:2006xi,Maiani:2017kyi,Wang:2022clw}. While in recent works~\cite{Esposito:2021vhu,Baru:2021ldu,Du:2021zzh}, it is pointed out that, from the viewpoint of effective ranges, the $X(3872)$ resonance should be viewed as an elementary core coupled with a sizeable $D\bar{D}^*$ component in the continuum. In fact, a new method was proposed in Ref.~\cite{Wang:2015rcz} to determine the short distance $c\bar{c}$ component of $X(3872)$ from its production in the semileptonic and nonleptonic $B_c$ decays.

Because of the high efficiencies and very good mass resolution in the construction of  the decay modes of the $X(3872)$ resonance, the open-charm decays and radiative transitions of the $X(3872)$ resonance have been investigated by the BESIII Collaboration~\cite{BESIII:2020nbj,Yuan:2022lxf}, and it was found that the experimental results, taking into account the model predictions~\cite{Ferretti:2014xqa,Braaten:2019gwc,Braaten:2020iye}, support that the $X(3872)$ resonance is more likely a molecule or a mixture of a molecule and a  charmonium, rather than a pure charmonium state.

Nevertheless, the interpretation of $X(3872)$ as a $D\bar{D}^*$ molecule  is the most popular, due to its proximity  to the $D\bar{D}^*$ threshold, which explains naturally the large isospin violating $J/\psi \rho^0$ decay mode~\cite{Gamermann:2009fv,Li:2012cs,Meng:2021kmi,Suzuki:2005ha,Liu:2006df,Gamermann:2009uq,Coito:2010if,Albaladejo:2015dsa,Wu:2021udi}. In addition, using an effective hadron theory based on the hidden-gauge Lagrangian, the $X(3872)$ resonance can be dynamically generated from the $S$-wave interaction of a pair of pseudoscalar and vector charmed mesons~\cite{Montana:2022inz}. While within the nonrelativistic effective field theory, the one boson exchange model and the molecular hypothesis of the $X(3872)$ resonance as a $D\bar{D}^*$ bound state, other $S$-wave hadronic molecules formed
by a pair of ground state charmed and anticharmed mesons are proposed~\cite{Liu:2008fh,Liu:2019stu,Ji:2022uie}. In Ref.~\cite{Wang:2022qxe}, based on the molecular nature of $X(3872)$, its charmless decays via intermediate meson loops were studied.

By studying the production of $X(3872)$ in $B$ and $B_s$ decays, more information of $X(3872)$ can be extracted~\cite{BaBar:2019hzd,CMS:2020eiw,Maiani:2020zhr,Ovsiannikova:2021hpn,Zhang:2022xqs,Yan:2022dnj}. In fact, it was shown that the nonleptonic weak decays of $B$ mesons can be useful tools to study hadronic resonances, some of which are subjects of intense debate about their nature~\cite{Oset:2016lyh}. In addition, those weak decays are also helpful to investigate final-state interactions and hence have the potential to shed further light on the nature of some puzzling hadrons~\cite{Liang:2014ama,Liang:2015twa,Xie:2018rqv,Liu:2020orv}. For example, in a recent work~\cite{Liu:2022zbd}, the weak decays of $B \to \bar{D}^{(\ast)}D_{s0}^{*}(2317)$ and $B \to \bar{D}^{(\ast)}D_{s1}(2460)$ are investigated by including the triangle diagrams where the $B$ meson first decays weakly into $\bar{D}^{(\ast)}D_{s}^{(\ast)}$ and $J/\psi K$($\eta_{c}K$), and then the $D_{s0}^{\ast}(2317)$ and $D_{s1}(2460)$ are dynamically generated by the final-state interactions of $D_{s}^{(\ast)}\eta$ and $D^{(\ast)}K$ via exchanges of $\eta$ and $D^{(\ast)}$ mesons. The obtained branching fraction of $B \to \bar{D}^{(\ast)}D_{s0}^{*}(2317)$ is in reasonable agreement with the experimental data.

In a recent work~\cite{CMS:2020eiw}, the CMS Collaboration observed the $B^0_s \to \phi X(3872)$ decay. Meanwhile, it was found that the ratio of $\mathcal{B}[B_s^0\to\phi X(3872)]/\mathcal{B}[B^0 \to K^0 X(3872)]$ is consistent with one, while the ratio of $\mathcal{B}[B_s^0\to\phi X(3872)]/\mathcal{B}[B^+ \to K^+ X(3872)]$ is two times smaller. This supports the previous measurement by the Belle Collaboration~\cite{Belle:2011vlx}:
\be 
R_K &=& \frac{\mathcal{B}[B^0\to K^0 X(3872)]}{\mathcal{B}[B^+\to K^+ X(3872)]} \nonumber \\ 
&=& 0.50 \pm 0.14(\text{stat.}) \pm 0.04(\text{syst.}). \label{equation:focus}
\ee 
This indicates a difference in the production dynamics of the $X(3872)$ resonance in $B^0$ and $B^+$ decays. 

Within the compact tetraquark picture that the $X(3872)$ resonance is a mixture  of four quark states $X_u = [cu][\bar{c}\bar{u}]$ and $X_d = [cd][\bar{c}\bar{d}]$, the above difference shown in Eq.~\eqref{equation:focus}  can be explained with certain phenomenological model parameters~\cite{Maiani:2020zhr}. Meanwhile,  the production rates  of the charged tetraquark states $X^{+} = [cu][\bar{c}\bar{d}]$ and $X^- = [cd][\bar{c}\bar{u}]$ are also predicted, which vary widely.

Before the experimental measurements of the $B^0 \to K^0 X(3872)$ and $B^+ \to K^+ X(3872)$ decays, a pioneering theoretical study of the exclusive production of the $X(3872)$ resonance in $B$ meson decays was performed in Ref.~\cite{Braaten:2004ai}. In that work, in the molecular picture where the $X(3872)$ resonance is a loosely bound $S$-wave state of the charmed mesons $D^0\bar{D}^{*0}$ or $\bar{D}^0 D^{*0}$~\cite{Braaten:2003he}, the ratio of the branching fractions of $B^0 \to X(3872) K^0$ and $B^+ \to X(3872) K^+$ is expressed in terms of the model parameters that parameterize the amplitudes of the $B \to \bar{D} D^*K$ decays. Based on the determined model parameters from the analysis of the experimental data, it was shown that the calculated branching fraction of the $B^0 \to K^0 X(3872)$ decay is suppressed by more than one order of magnitude compared to that of the $B^+ \to K^+ X(3872)$ decay.

Indeed, the quoted branching fractions of $B^0 \to K^0 X(3872)$ and $B^+ \to K^+ X(3872)$ decays by the review of particle physics (RPP) are~\cite{ParticleDataGroup:2020ssz}
\begin{eqnarray}
{\cal B} [B^0 \to K^0 X(3872)] &=& (1.1 \pm 0.4) \times 10^{-4}, \\
{\cal B} [B^+ \to K^+ X(3872)] &=& (2.1 \pm 0.7) \times 10^{-4},
\end{eqnarray}
from which one can deduce $R_K = 0.52 \pm 0.26$, which is in agreement with the Belle result shown in Eq.~\eqref{equation:focus} within uncertainties.

In this work, based on the molecular picture where the  $X(3872)$ resonance is strongly coupled to the $D\bar{D}^*$ channels,~\footnote{For simplicity, in the rest of this paper,  the  charge conjugate states are always implied for $D\bar{D}^*$, $D^+D^{\ast-}$, and $D^0\bar{D}^{\ast0}$ unless otherwise stated.} we study the decays of $B^0 \to K^0 X(3872)$ and $B^+ \to K^+ X(3872)$. By considering the long-distance production of the $X(3872)$ resonance from the rescattering of $D$ and $\bar{D}^*$ and the short-distance production of the $X(3872)$ resonance from the $c\bar{c}$ pair, we investigate the difference of the above two decays. Furthermore, with the same reaction mechanisms and model parameters, the decays of $B_s^0 \to \eta (\eta') X(3872)$ are  evaluated.

This article is organized as follows. In Sec.~\ref{sec:formalism}, we present the theoretical formalism for the production of $X(3872)$ resonance in the $B$ meson decays, and in Sec.~\ref{sec:results}, we show our theoretical numerical results and discussions, followed by a short summary in the last section.

%%%%%%%%%%%%%%%%%%%%%%%%%%%%%%%%%%%%%%%%%%%%%%%%%%%%%%%%%%%%%%%%%%%%%%%%%%

\section{Theoretical formalism} \label{sec:formalism}

In this section, we present the two kinds of reaction mechanisms for the $B^0 \to K^0 X(3872)$ and $B^+ \to K^+ X(3872)$ decays. One is that the production of $X(3872)$  is mainly from the short-distance dynamics, namely the hadronization of the $c\bar{c}$ pair. The other one is the long-distance contribution where the $X(3872)$ resonance was produced from the rescattering of the charmed meson pair $D$ and $\bar{D}^*$.

\subsection{Decays of $B^{0} \to K^{0} X(3872)$ and $B^{+} \to K^{+} X(3872)$}

\begin{figure}[htbp]
\centering
\subfigure[]{
\includegraphics[scale=0.6]{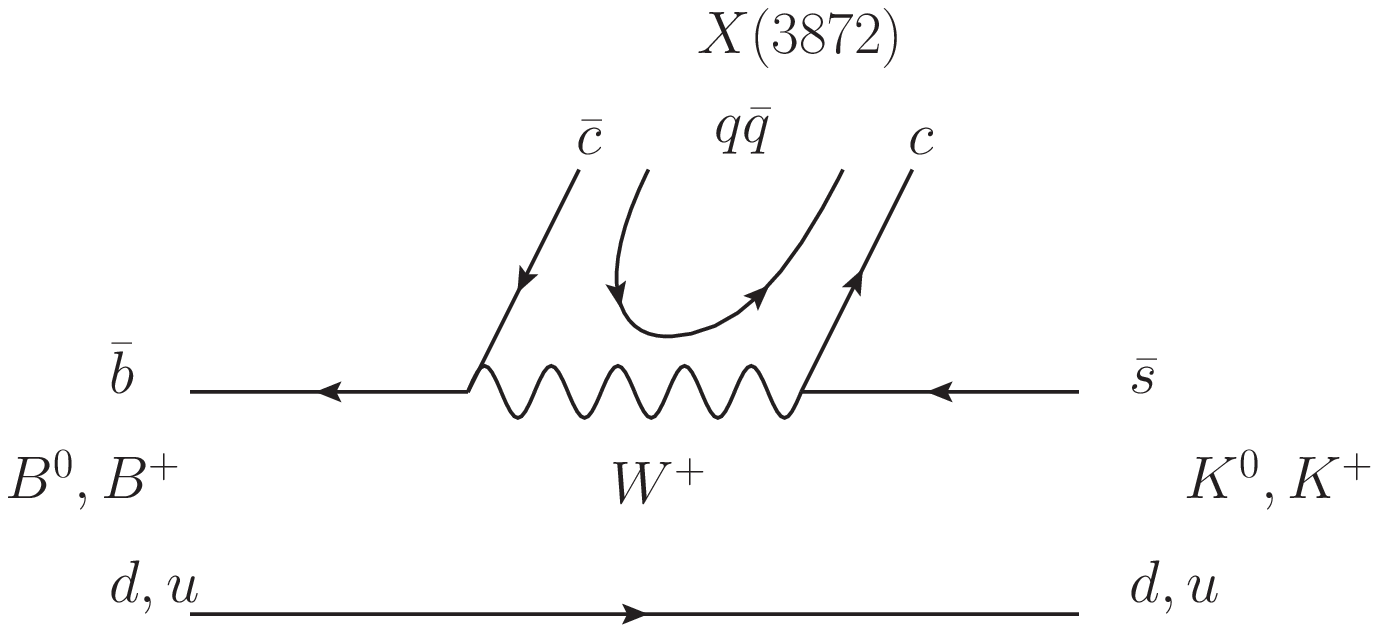}}
\subfigure[]{\includegraphics[scale=0.6]{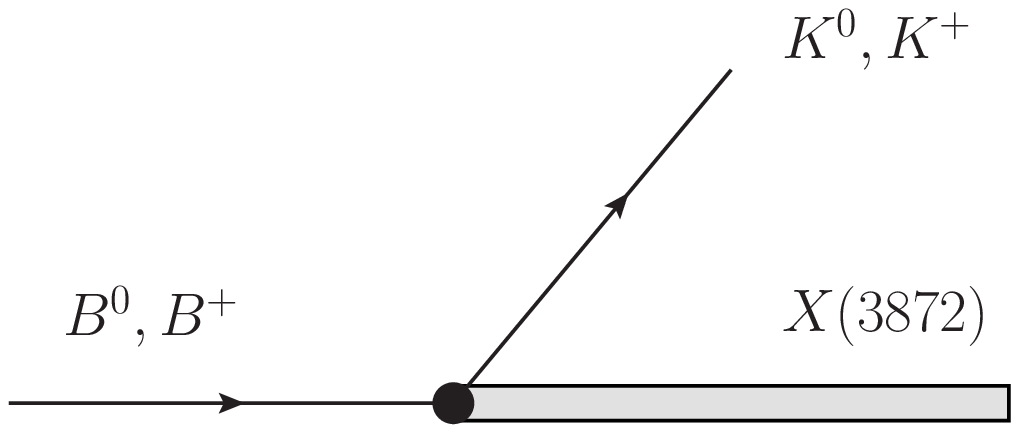}}
\caption{Quark-level and hadron-evel Feynman diagrams of the $B^0 \to K^0 X(3872)$ and $B^+ \to K^+ X(3872)$ decays. The $c\bar{c}$ pair is hadronized into the $X(3872)$ resonance with or without a pair of sea quarks.} \label{fig:B0plus}
\end{figure}

For the $B^{0} \to K^{0} X(3872)$ and $B^{+} \to K^{+} X(3872)$ decays, the $X(3872)$ resonance can be produced directly from the hadronization of the $c\bar{c}$ pair via the $W$-internal exchange diagram of the weak decay, as shown in Fig.~\ref{fig:B0plus} (a). In this way, the $X(3872)$ resonance is produced mainly from the short-distance dynamics, whose contribution to the $B^0 \to K^0 X(3872)$ and $B^+ \to K^+ X(3872)$ decays are expected to be the same, up to the small differences between the particle masses involved in the above two decays.

For the diagrams shown in Fig.~\ref{fig:B0plus} (b), the corresponding decay amplitudes can be written as
\be 
t_1(B^0) &=& g_{BKX} p_{B^0}^{\mu}\epsilon_{\mu}^{\ast}(X), \\
t_1(B^+) &=& g_{BKX} p_{B^+}^{\mu}\epsilon_{\mu}^{\ast}(X),
\ee
where $p_{B^0}$ and $p_{B^+}$ are the four-momenta of the $B^0$ and $B^+$ mesons, respectively, and $\epsilon(X)$ is the polarization vector of the $X(3872)$ resonance. Note that  the same effective coupling strength $g_{BKX}$ is taken to be the same.

As seen in Fig.~\ref{fig:B0plus} (a), there are also contributions from rescattering of $D$ and $\bar{D}^*$ that are hadronized from the $c\bar{c}$ pair. However, these contributions can be easily absorbed into the coupling strength of $g_{BKX}$. 

\begin{figure}[htbp]
\centering
\subfigure[]{
\includegraphics[scale=0.5]{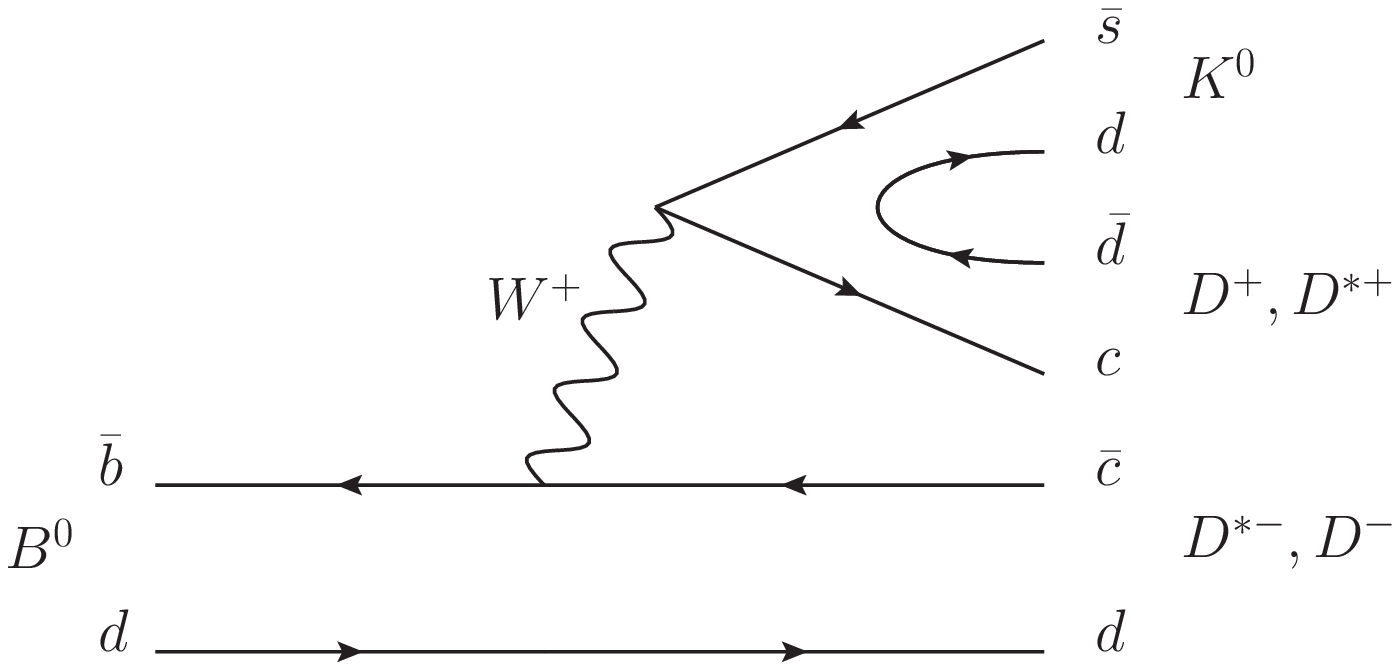}}
\subfigure[]{
\includegraphics[scale=0.5]{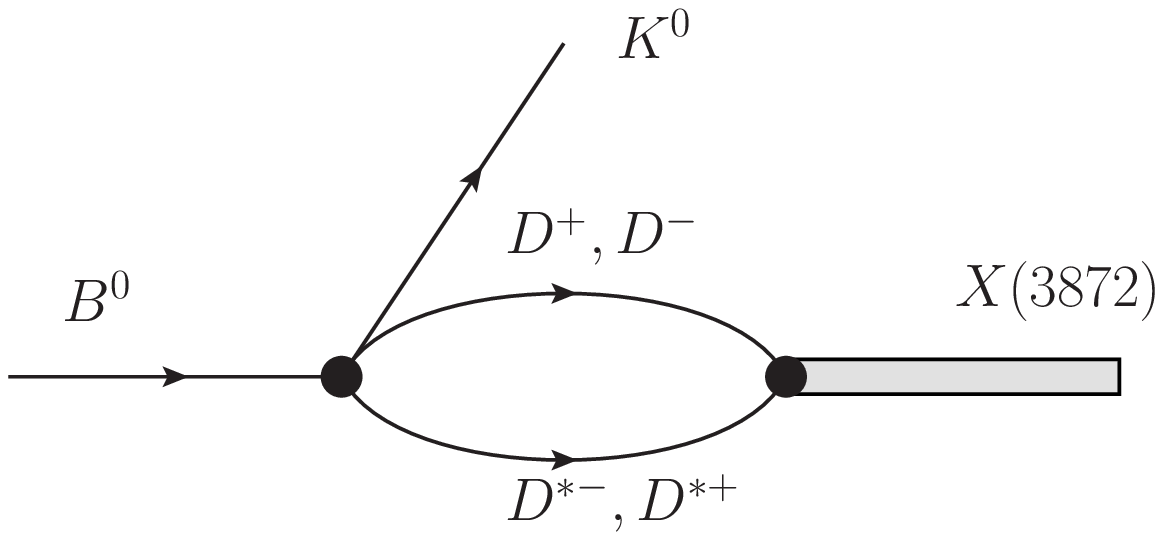}}
\caption{Quark-level and hadron-level Feynman diagrams of the $B^0 \to K^0 X(3872)$ decay.} \label{fig:B0}
\end{figure}

On the other hand, the $B^0 \to K^0 X(3872)$ decay can also proceed via the following processes: in the first step the anti-bottom quark in the $B^0$ meson turns into an anti-charm quark and a $\bar{s} c$ pair via the $W$-external emission diagram, which is, in general, the dominant term of the weak decays~\cite{Chau:1987tk}. The next step consists in introducing a pair of sea quarks $d\bar{d}$ with the quantum numbers of the vacuum, to form a $K^0$ and a $D^+$ (or $D^{*+}$) with the $\bar{s}c$ pair. While the $\bar{c}$ and $d$ quark from the $B^0$ meson will be hadronized into a $D^{*-}$ (or $D^-$) meson. These above processes are shown in Fig.~\ref{fig:B0} (a). Finally, the final-state interactions of the $D^+ D^{*-}$ and $D^{*+}D^-$ will lead to the production of the $X(3872)$ resonance, as shown in Fig.~\ref{fig:B0} (b).

\begin{figure}[htbp]
\centering
\subfigure[]{
\includegraphics[scale=0.5]{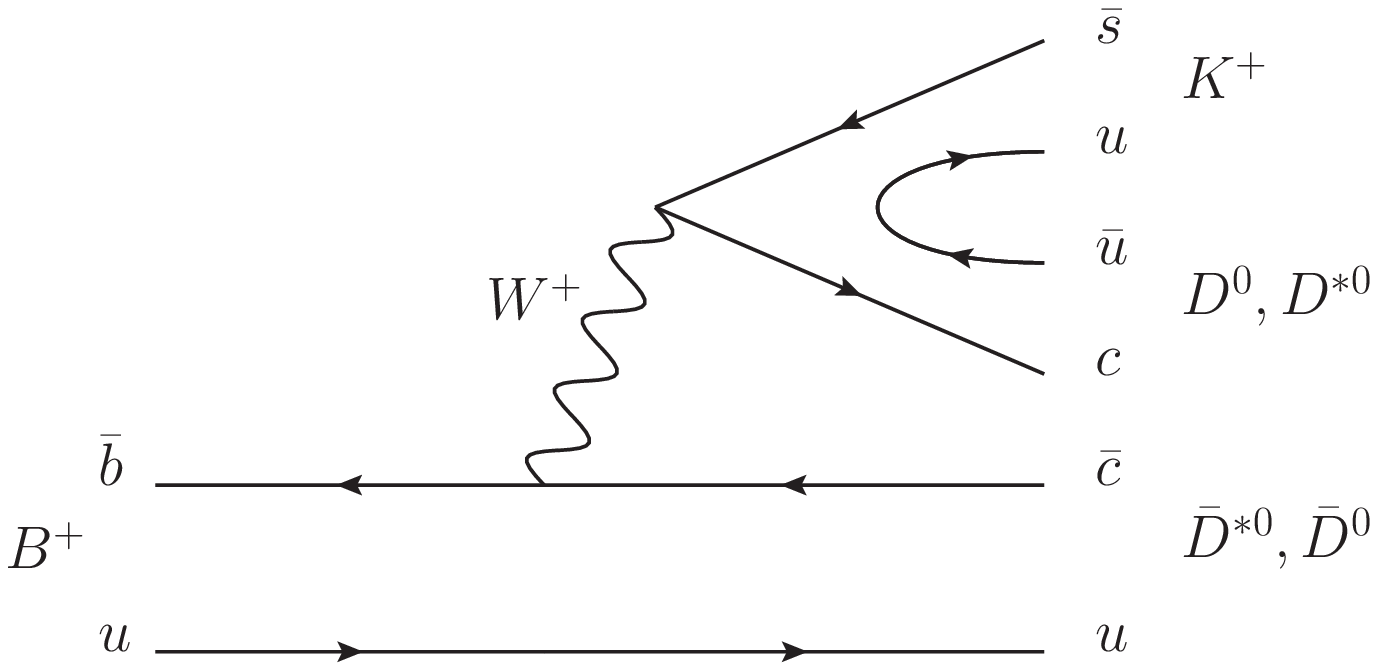}}
\subfigure[]{
\includegraphics[scale=0.5]{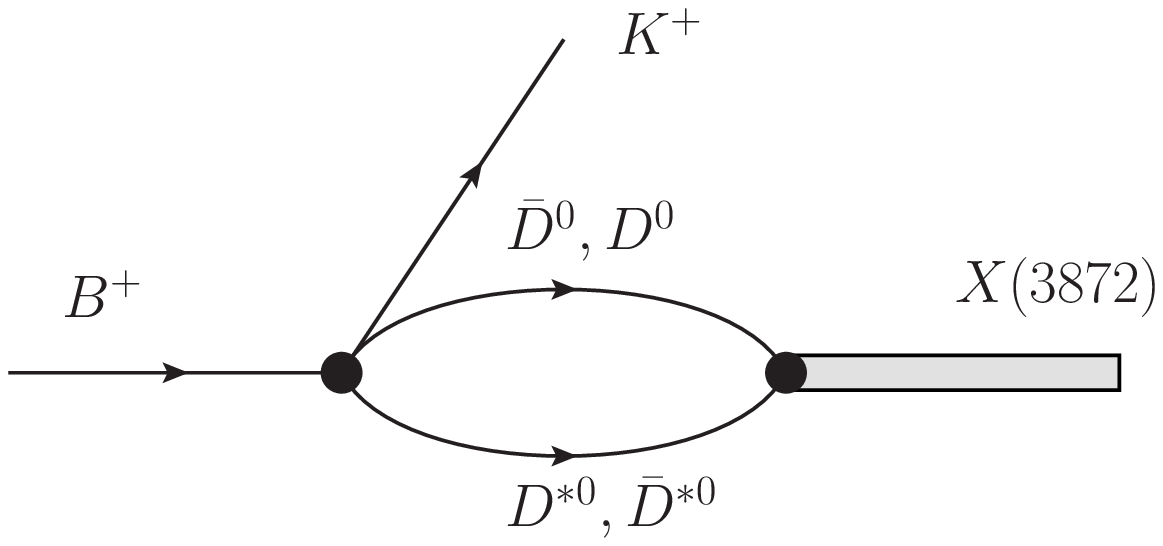}}
\caption{Quark-level and hadron-level Feynman diagrams of the $B^+\to K^+ X(3872)$ decay.} \label{fig:Bplus}
\end{figure}

The $B^+ \to K^+ X(3872)$ decay can also occur in the same way, as shown in Fig.~\ref{fig:Bplus}. In this case, to produce the $X(3872)$ resonance in the final state, one has to produce the $\bar{D}^0 D^{*0}$ or $D^0 \bar{D}^{*0}$ states in the hadronization process. Then the $X(3872)$ resonance will be produced by the final state interactions of charmed $D^0$ and $\bar{D}^{*0}$ mesons.

For the $B \to K D \bar{D}^*$ vertex, it is in $P$-wave because of angular momentum conservation. Thus the  amplitudes of Fig.~\ref{fig:B0} (b) and Fig.~\ref{fig:Bplus} (b) for the $B^0 \to K^0 X(3872)$ and $B^+ \to K^+ X(3872)$ decays can be written as
\be 
    t_2(B^0) &=& V_P g_{X(3872)D^+\bar{D}^{*-}} p_{B^0}^{\mu}G_{D^+D^{\ast-}}\epsilon_{\mu}^{\ast}(X), \\
    t_2(B^+) &=& V_P g_{X(3872)D^0\bar{D}^{*0}} p_{B^+}^{\mu}G_{D^0\bar{D}^{\ast0}}\epsilon_{\mu}^{\ast}(X),
\ee 
where $V_P$ is a global production factor, containing the dynamics which is common to the $B^0 \to K^0 X(3872)$ and $B^+ \to K^+ X(3872)$ decays. While $g_{X(3872)D^+\bar{D}^{*-}}$ and $g_{X(3872)D^0\bar{D}^{*0}}$ are the coupling constants of the $X(3872)$ resonance to the charged and neutral $D\bar{D}^*$ channels. Besides, in the above equations, $G$ is the loop function of $D$ and $\bar{D}^*$, which reads~\cite{Gamermann:2007fi,Gamermann:2009fv,Montana:2022inz},
\be 
    G &=& \frac{1}{16\pi^2}\left( \alpha + \log\frac{m_1^2}{\mu^2}+\frac{m_2^2-m_1^2+s}{2s}\log\frac{m_2^2}{m_1^2}\right.\notag\\
    &&\left.+\frac{p}{\sqrt{s}}\left( \log\frac{s-m_2^2+m_1^2+2p\sqrt{s}}{-s+m_2^2-m_1^2+2p\sqrt{s}}\right.\right.\notag\\
    &&\left.\left.+\log\frac{s+m_2^2-m_1^2+2p\sqrt{s}}{-s-m_2^2+m_1^2+2p\sqrt{s}}\right)\right), \label{eq:Gloop}
\ee
where $m_1$ and $m_2$ are the masses of $D$ and $\bar{D}^*$ mesons, respectively. The subtraction constant $\alpha$ at the regularization scale $\mu$ is a free parameter. As done in Refs.~\cite{Montana:2022inz}, we fix $\mu = 1000$ MeV. Note that $\mu$ and $\alpha$ are not independent, this justifies setting the energy scalre $\mu$ at a fixed value and fitting just $\alpha$ to data. In the following, the value of $\alpha$ will be determined by the experimental branching fractions of $B^0 \to K^0 X(3872)$ and $B^+ \to K^+ X(3872)$.

In Eq.~\eqref{eq:Gloop}, $p$ is the three momentum of the two mesons in the center mass frame of $X(3872)$, which is 
\be
p = \frac{\lambda^{1/2}(s,m^2_1,m^2_2)}{2\sqrt{s}},
\ee
with the K\"allen triangle function $\lambda(x,y,z) = x^2 + y^2 + z^2 - 2xy - 2yz - 2zx$,  and $s = M^2_{X(3872)}$. 

The two amplitudes $t_1$ and $t_2$ from the two different reaction mechanisms play a vital role in the decays of $B^{0(+)}\to K^{0(+)}X(3872)$. The total decay widths read~\cite{ParticleDataGroup:2020ssz}
\be 
\Gamma_{B^0\to K^0X(3872)} &=& \frac{1}{8\pi}\frac{|\bold{q}|^3}{M^2_X} \times \notag\\
    && \left( g_{BKX}^2 + g^2_{B^0} |G_{D^+D^{\ast-}}|^2 \right),
    \label{equation:11} \\
    && = \Gamma^{\rm I}_{B^0} + \Gamma^{\rm II}_{B^0}, \notag \\
\Gamma_{B^+\to K^+X(3872)} &=& \frac{1}{8\pi}\frac{|\bold{q}|^3}{M^2_{X}} \times \notag\\
    && \left( g_{BKX}^2 + g^2_{B^+} |G_{D^0D^{\ast0}}|^2 \right),
    \label{equation:12} \\
    && = \Gamma^{\rm I}_{B^+} + \Gamma^{\rm II}_{B^+}, \notag
\ee
where $M_X$ stands for the mass of the $X(3872)$ resonance, and $g_{B^0} = V_P g_{X(3872)D^+\bar{D}^{*-}}$ and $g_{B^+} = V_P g_{X(3872)D^0\bar{D}^{*0}}$. $\Gamma^{\rm I}_{B^0/B^+}$ and $\Gamma^{\rm II}_{B^0/B^+}$ represent the contributions from $t_1$ and $t_2$,~\footnote{The interference terms are not considered here, since, in general, these contributions are smaller than the two main processes. Furthermore, including such contributions, more parameters are needed, and one can not determine or constrain these parameters at present. Hence, we will investigate the interference terms in future works when more experimental data become available.} respectively. While the three-momentum $\bold{q}$ is
\be
|\bold{q}| = \frac{\lambda^{1/2}(M^2_{B^0/B^+},m^2_{K^0/K^+},M^2_X)}{2M_{B^0/B^+}},
\ee
where $M_{B^0/B^+}$ and $m_{K^0/K^+}$ are the masses of $B^{0(+)}$ and $K^{0(+)}$, respectively. In this work, we take $M_{B^0} = 5279.65$ MeV, $M_{B^+} = 5279.34$ MeV, $m_{K^0} = 497.611$ MeV, and $m_{K^+} = 493.677$ MeV.

\subsection{Decays of $B_s^0 \to \eta (\eta') X(3872)$}

\begin{figure}[htbp]
\centering
\subfigure[]{
\includegraphics[scale=0.6]{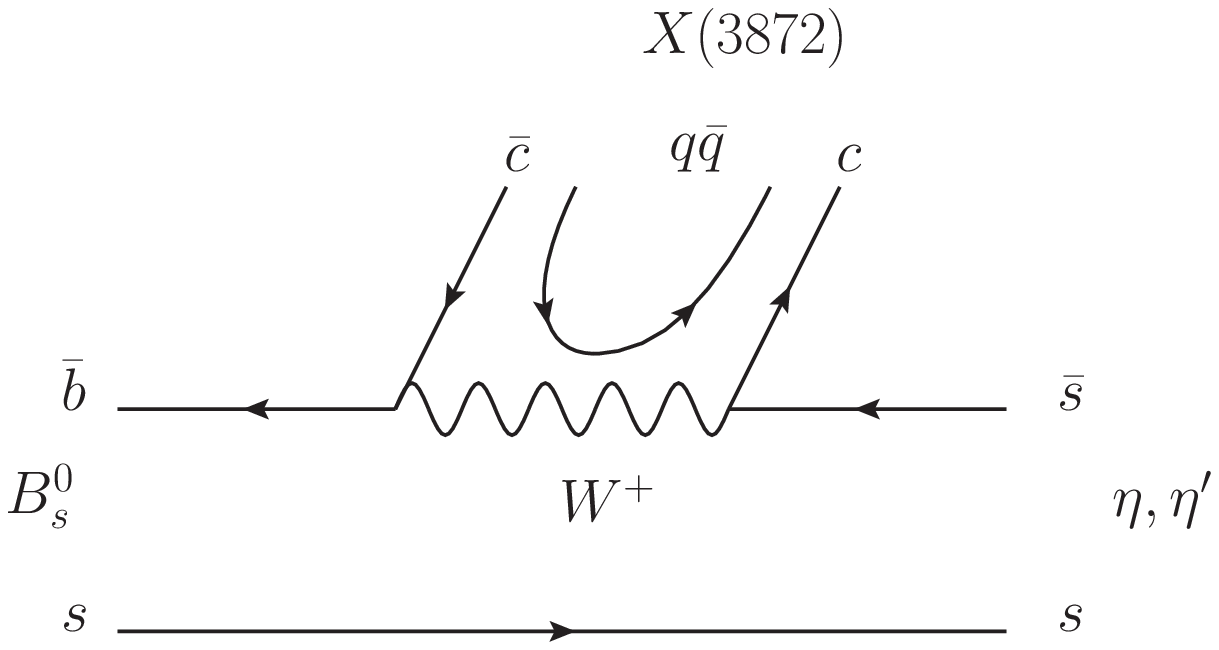}}
\subfigure[]{
\includegraphics[scale=0.6]{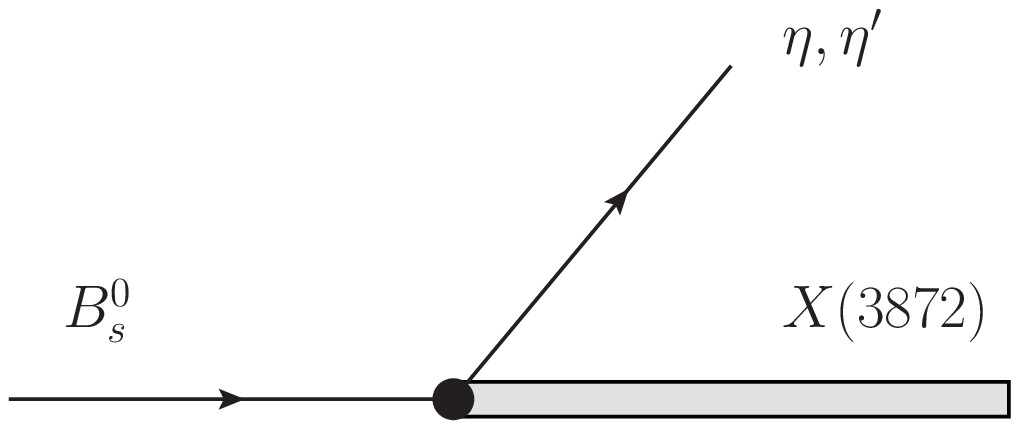}}
\caption{Quark-level and hadron-level Feynman diagrams of the $B_s^0 \to \eta(\eta')X(3872)$ decays.} \label{fig:Bsdecay}
\end{figure}

Based on the mechanisms for the $B^{(0,+)} \to K^{(0,+)} X(3872)$ decays, one can also investigate the processes of $B_s^0\to \eta(\eta') X(3872)$, where the $X(3872)$ resonance is produced from the hadronization of the $c\bar{c}$ pair. The quark and hatron level diagrams ofthe  $B_s^0\to \eta (\eta') X(3872)$ decay are shown in Fig.~\ref{fig:Bsdecay}. Considering $SU(3)$ flavor symmetry, the corresponding amplitude is
\be 
    t(B^0_s) = g_{BKX} p_{B_s^0}^{\mu} \epsilon_{\mu}^{\ast}(X).
\ee 

Then, the partial decay widths for the processes $B^0_s \to \eta X(3872)$ and $B^0_s \to \eta' X(3872)$ are given by
\be 
    && \Gamma_{B_s^0 \to \eta(\eta') X(3872)} = \frac{g^2_{BKX}f_{\eta/\eta'}}{8\pi}\frac{|\bold{p}_{\eta/\eta'}|^3}{M^2_X}, \\
  && |\bold{p}_{\eta/\eta'}| = \frac{\lambda^{1/2}(M^2_{B^0_s},m^2_{\eta/\eta'},M^2_X)}{2M_{B^0_s}}.  
\ee
We take $M_{B^0_s} = 5366.91$ MeV, $m_{\eta} = 547.86$ MeV, and $m_{\eta'} = 957.78$ MeV. In addition, we also need  the coefficients $f_{\eta}$ and $f_{\eta'}$ since the $\eta$ and $\eta'$ mesons are the mixture of flavor eigen-states $\eta_1$ and $\eta_8$, and we take $f_{\eta}=1/3$ and $f_{\eta'}= 2/3$ as in Refs.~\cite{Bramon:1992kr,Miyahara:2016yyh,Guo:2005wp,Gu:2018swy,Gao:2022xqz}. The flavor wave functions of the $\eta$ and $\eta'$ mesons are:
\be 
 \eta  & = & \frac{1}{\sqrt{3}} (u\bar{u} + d\bar{d} - s\bar{s}), \\
 \eta' & = & \frac{1}{\sqrt{6}} (u\bar{u} + d\bar{d} + 2 s\bar{s}).
\ee 

Note that in Ref.~\cite{Montana:2022inz}, it is found that the $X(3872)$ resonance can be dynamically generated from $D\bar{D}^*$ and $D_s\bar{D}^*_s$ coupled channels in $S$-wave interactions, and the $X(3872)$ resonance has a sizeable coupling to the $D_s\bar{D}^*_s$ channel. Thus, for the $B^0_s \to \eta (\eta') X(3872)$ decays there should be also long-distance contributions from $D_s$ and $\bar{D}^*_s$ rescattering. However, since the $D_s\bar{D}^*_s$ mass threshold  is much higher than the  $X(3872)$ mass, the loop function $G_{D_s\bar{D}^*_s}$ will be very small at the  $X(3872)$ mass, thus it is expected that the contributions from these diagrams should be tiny and we ignore them here. 

%%%%%%%%%%%%%%%%%%%%%%%%%%%%%%%%%%%%%%%%%%%%%%%%%%%%%%%%%%%%%%%%%%%%%%%%%%

\section{Numerical results and discussions}\label{sec:results}

To calculate the branching fractions of ${\cal B}[B^0 \to K^0 X(3872)]$ and ${\cal B}[B^+ \to K^+ X(3872)]$, we need to know the coupling constants $g_{X(3872)D^0\bar{D}^{*0}}$ and $g_{X(3872) D^+ \bar{D}^{*-}}$. Here, we  take them as the same, as expected in the molecular picture  of the $X(3872)$ resonance~\cite{Albaladejo:2015dsa,Guo:2014hqa,Ji:2022uie}, and then absorb them into $g_{B^0}$ and $g_{B^+}$. In doing this, we can reduce the number of model parameters. In addition, we also need to determine the other two unknown couplings $g_{BKX}$, $V_P$, and the subtraction constant $\alpha$.

One can find that the equations (\ref{equation:11}) and (\ref{equation:12}) mainly differ in the loop functions $G_{D^0\bar{D}^{*0}}$ and $G_{D^+\bar{D}^{*-}}$. In other words, the difference between the two decays $B^0 \to K^0 X(3872)$ and $B^+ \to K^+ X(3872)$ originates mainly from the different loop functions $G_{D^+D^{\ast-}}$ and $G_{D^0\bar{D}^{\ast0}}$.

Note that in Ref.~\cite{Wu:2021udi}, a larger coupling of the $X(3872)$ resonance to the neutral $D^0\bar{D}^{*0}$ channel is used. It is found that the difference coupling strengths of the $X(3872)$ resonance to its charged component $D^+\bar{D}^{*-}$ and neutral component $D^0\bar{D}^{*0}$ is one source of the strong isospin violation decays of $X(3872) \to J/\psi \rho^0$ and $X(3872) \to J/\psi \omega$. In fact, a larger value of $g_{X(3872)D^0\bar{D}^{*0}}$ is also welcome in this work.

\begin{figure}[htbp]
\centering
\includegraphics[scale=0.35]{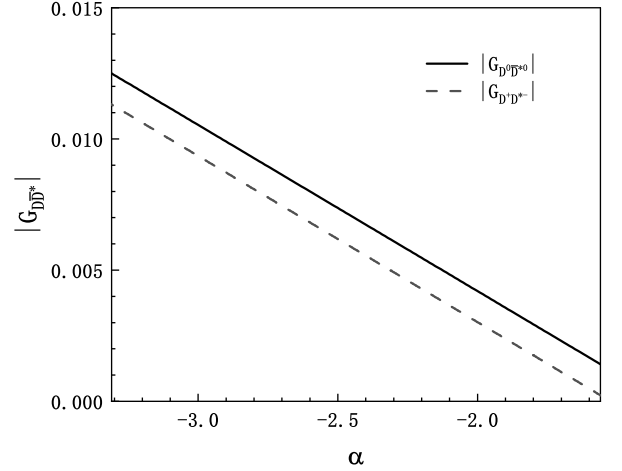}
\caption{Absolute values of $G_{D^+D^{\ast-}}$ and $G_{D^0\bar{D}^{\ast0}}$ as a function of the subtraction constant $\alpha$.}
\label{fig:Gloop}
\end{figure}

In Fig.~\ref{fig:Gloop} we show the numerical results for the loop functions $G_{D^+D^{\ast-}}$ and $G_{D^0\bar{D}^{\ast0}}$ with $M_{X(3872)} = 3871.65$ MeV, $m_{D^0} = 1864.84$ MeV, $m_{D^{*0}} = 2006.85$ MeV, $m_{D^+} = 1869.66$ MeV, $m_{D^{*-}} = 2010.26$ MeV. It is worthy to mention that, with the above values, the $X(3872)$  is located below the mass thresholds of $D^0\bar{D}^{*0}$ and $D^+\bar{D}^{*-}$. Hence, the loop functions $G_{D^+D^{\ast-}}$ and $G_{D^0\bar{D}^{\ast0}}$ are real and negative.

From Fig.~\ref{fig:Gloop}, one can see that the absolute values of $G_{D^0\bar{D}^{\ast0}}$ are larger than those of $G_{D^+D^{\ast-}}$. Therefore, it provides a natural explanation for the larger branching fraction of $B^+ \to K^+ X(3872)$ with the loop function $G_{D^0\bar{D}^{\ast0}}$ involved, where the $X(3872)$ resonance is produced by the rescattering of $D^0$ and $\bar{D}^{*0}$. In addition, as pointed out in Ref.~\cite{Wu:2021udi}, the coupling of $X(3872)$ to the neutral $D^0\bar{D}^{*0}$ channel is larger than the one to the charged $D^+\bar{D}^{*-}$ channel, which will enhance the production of $X(3872)$ in the $B^+ \to K^+ X(3872)$ decay.

To minimize the number of free parameters, we  fix the subtraction constant $\alpha$ at certain values,~\footnote{The natural value of the subtraction constant $\alpha$ is around $-2$ with $\mu$  about $1000$ MeV. In this work, it is found that one can reproduce the branching fractions of  ${\cal B}[B^0 \to K^0 X(3872)]$ and ${\cal B}[B^+ \to K^+ X(3872)]$ by varying the value of $\alpha$ in the range of $-2.09 < \alpha < -1.43$.} and then determine $g_{BKX}$ and $g_B = g_{B^0} = g_{B^+}$ by fitting them to the experimental branching fractions of the ${\cal B}[B^0 \to K^0 X(3872)]$ and ${\cal B}[B^+ \to K^+ X(3872)]$ decays. In fact, one can only determine their absolute values. Therefore, in this work, we take them as real and positive. The obtained results are listed in Table~\ref{tab:results}. The uncertainties of these fitted parameters are propagated from the uncertainties of the branching fractions ${\cal B}[B^0 \to K^0 X(3872)]$ and ${\cal B}[B^+ \to K^+ X(3872)]$.

In Ref.~\cite{Montana:2022inz}, the $X(3872)$ resonance is  generated  using the cutoff regularization for the $D\bar{D}^*$ loop functions with a cutoff  of 567 MeV, which corresponds to a subtraction constant $\alpha = -1.91$ for the $D\bar{D}^*$ channels with $\mu = 1000$ MeV.

\begin{table*}[htbp]
    \renewcommand\arraystretch{1.25}
    \caption{Fitted parameters of $g_{BKX}$ and $g_B$, the ratios of $R_1 = \Gamma_{B^0}^{\rm II}/\Gamma^{\rm I}_{B^0}$ and $R_2 = \Gamma_{B^+}^{\rm II}/\Gamma^{\rm I}_{B^+}$, and the so-obtained branching fractions of $B^0_s \to \eta X(3872)$ and $B^0_s \to \eta' X(3872)$ decays.}
    \center
    \begin{tabular}{ccccccccc}\hline\hline
    \multirow{2}{*}{$\alpha$} & \multicolumn{1}{c}{$|G_{D^+D^{\ast-}}|^2$} &     \multicolumn{1}{c}{$|G_{D^0D^{\ast0}}|^2$} &    \multirow{2}{*}{$g_{BKX} ~(\times10^{-8})$} &    \multirow{2}{*}{$g_B~(\times10^{-5})$} &    \multirow{2}{*}{$R_1$} &    \multirow{2}{*}{$R_2$}~~~~& \multicolumn{2}{c}{Theoretical branching factions $(\times 10^{-5})$}  \\  \cline{2-3} \cline{8-9}
    \multirow{2}{*}{ }& $(\times10^{-5})$ & $(\times10^{-5})$ &\multirow{2}{*}{ }&\multirow{2}{*}{ }& & &~~~~~~~~$B^0_s \to \eta X(3872)$ & $B^0_s \to \eta' X(3872)$  \\ \hline
$-1.86$ & $0.45$ & $1.10$ & $7.45\pm1.77
$ & $3.79\pm0.90
$ & $1.18\pm0.79$ & $2.85\pm1.92$ & $1.94\pm0.92$ & $2.16\pm1.02$ \\ 
     $-1.91$ & $0.60$ & $1.33$ & $6.61\pm1.57
$ & $3.59\pm0.85
$ & $1.78\pm1.19$ & $3.90\pm2.62$ & $1.52\pm0.72$ & $1.70\pm0.81$ \\ 
     $-1.96$ & $0.77$ & $1.56$ & $5.62\pm1.33
$ & $3.42\pm0.81
$ & $2.83\pm1.90$ & $5.77\pm3.87$ & $1.10\pm0.52$ & $1.23\pm0.58$ \\ 
     $-2.01$ & $0.95$ & $1.82$ & $4.42\pm1.05 
$ & $3.27\pm0.78
$ & $5.20\pm3.49$ & $9.94\pm6.68$ & $0.68\pm0.32$ & $0.76\pm0.36$ \\ 
     $-2.06$ & $1.16$ & $2.10$ & $2.72\pm0.65
$ & $3.14\pm0.74
$ & $15.35\pm10.30$ & $27.86\pm18.70$ & $0.26\pm0.12$ & $0.29\pm0.14$ \\ 
     $-2.08$ & $1.25$ & $2.22$ & $1.60\pm0.38
$ & $3.09\pm0.63
$ & $46.31\pm31.11$ & $82.53\pm55.45$ & $0.09\pm0.04$ & $0.10\pm0.05$ \\ \hline\hline
    \end{tabular}
    \label{tab:results}
\end{table*}

Based on the fitted parameters $g_{BKX}$ and $g_B$, we can calculate the ratio of the two contributions to the $B^0 \to K^0 X(3872)$ and $B^+ \to K^+ X(3872)$ decays:
\begin{eqnarray}
R_1 = \frac{\Gamma_{B^0}^{\rm II}}{\Gamma^{\rm I}_{B^0}}, ~~~~~~~
R_2 = \frac{\Gamma_{B^+}^{\rm II}}{\Gamma^{\rm I}_{B^+}}.
\end{eqnarray}
The theoretical results are also shown in Table~\ref{tab:results}. On can see that the long-distance contributions is dominant for the production of the $X(3872)$ resonance in the $B^0$ and $B^+$ decays, particularly the latter, for most of the parameter space explored.

Next, we turn to the $B^0_s \to \eta X(3872)$ and $B^0_s \to \eta' X(3872)$ decays. Based on the fitted parameter of $g_{BKX}$, we can calculate the partial decay widths of $B_s^0\to \eta(\eta')X(3872)$. The theoretical predictions are also listed in Table~\ref{tab:results}. It is found that their fractions are  smaller than those of the ${\cal B}[B^0 \to K^0 X(3872)]$ and ${\cal B}[B^+ \to K^+ X(3872)]$ decays by one order of magnitude. We hope that these predictions can be tested in future experiments. In addition, future experimental measurements of the $B_s^0 \to \eta X(3872)$ and $B_s^0 \to \eta' X(3872)$ decays will help us to constrain the value of the subtraction constant $\alpha$ in the loop functions of the charmed $D$ and $\bar{D}^*$ mesons.

One might think that the decay formalism proposed here can also be used for the $B^0_s \to \phi X(3872)$ decay. However, it is known that the $\phi$ meson is a vector state with spin-parity quantum numbers $J^P = 1^-$, then the vertex structure of $B^0_s \to \phi X(3872)$ is different from those of the kaon production. One should consider other contributions to the $B^0_s \to \phi X(3872)$ decay, and we will explore such a process in further works.

%%%%%%%%%%%%%%%%%%%%%%%%%%%%%%%%%%%%%%%%%%%%%%%%%%%%%%%%%%%%%%%%%%%%%%%%%%

\section{Summary} \label{sec:summary}

In the present work, we have studied the $X(3872)$ resonance production in the $B$ meson decays. By considering the long-distance component of the $X(3872)$ resonance, the difference between the $B^0 \to K^0 X(3872)$ and $B^+ \to K^+ X(3872)$ decays can be naturally explained. The enhancement of the $B^+ \to K^+ X(3872)$ decay over the $B^0 \to K^0 X(3872)$ decay is a nontrivial prediction  in the molecular picture of the $X(3872)$ resonance where it has strong couplings to the $D\bar{D}^*$ channel and is generated from the rescattering of the charmed $D$ and $\bar{D}^*$ mesons. The mass difference between the neutral and charged charmed mesons leads to different loop functions at the  $X(3872)$ mass, which provides the main factor accounting for the difference between the $B^0 \to K^0 X(3872)$ and $B^+ \to K^+ X(3872)$ decays. The present results provide further support that the $X(3872)$ resonance is not a pure charmonium state and contains a large $D\bar{D}^*$ component in its wave function.

Based on the fitted parameters from the branching fractions of the $B^0 \to K^0 X(3872)$ and $B^+ \to K^+ X(3872)$ decays, we also calculated the partial decay widths of $B_s^0\to \eta X(3872)$ $B_s^0\to \eta' X(3872)$, which are not yet measured. It is expected that these results can be tested in the future by the Belle II and LHCb collaborations. 

\section*{ACKNOWLEDGEMENT}

This work is partly supported by the National Natural Science Foundation of China under Grant Nos. ~12075288,~11735003,~11961141012,~12035007,~11975041, and 11961141004, the Youth Innovation Promotion Association CAS, Guangdong Provincial funding with Grant No.~2019QN01X172, Science and Technology Program of Guangzhou No.~2019050001. Q.W. is also supported by the NSFC and the Deutsche Forschungsgemeinschaft (DFG, German Research Foundation) through the funds provided to the Sino-German Collaborative Research Center TRR110 ``Symmetries and the Emergence of Structure in QCD"
(NSFC Grant No. 12070131001, DFG Project-ID 196253076-TRR 110).
%%%%%%%%%%%%%%%%%%%%%%%%%%%%%%%%%%%%%%%%%%%%%%%%%%%%%%%%%%%%%%%%%%%%%%%%%%

%\bibliographystyle{unsrt}
%\bibliography{spectral} 

\end{document}